\documentclass[twocolumn,showpacs,superscriptaddress,preprintnumbers,prl]{revtex4}

\usepackage{graphicx}
\usepackage{dcolumn}
\usepackage{bm}

\begin{document}
\title{Magnetic and Electronic Raman Scattering at the Nodal Spin-Density-Wave 
Transition in BaFe$_2$As$_2$}
\author{S. Sugai}
\affiliation{Department of Physics, Faculty of Science, Nagoya University, 
Furo-cho, Chikusa-ku, Nagoya 464-8602, Japan}
\affiliation{Venture Business Laboratory, 
Nagoya University, Furo-cho, Chikusa-ku, Nagoya 464-8601, Japan}
\affiliation{TRIP, Japan Science and Technology Agency (JST), Chiyoda, 
Tokyo 102-0075, Japan}
\author{Y. Mizuno}
\affiliation{Department of Physics, Faculty of Science, Nagoya University, 
Furo-cho, Chikusa-ku, Nagoya 464-8602, Japan}
\affiliation{TRIP, Japan Science and Technology Agency (JST), Chiyoda, 
Tokyo 102-0075, Japan}
\author{R. Watanabe}
\affiliation{Department of Crystalline Materials Science, Nagoya University, 
Furo-cho, Chikusa-ku, Nagoya 464-8603, Japan}
\affiliation{TRIP, Japan Science and Technology Agency (JST), Chiyoda, 
Tokyo 102-0075, Japan}
\author{T. Kawaguchi}
\affiliation{Department of Crystalline Materials Science, Nagoya University, 
Furo-cho, Chikusa-ku, Nagoya 464-8603, Japan}
\affiliation{TRIP, Japan Science and Technology Agency (JST), Chiyoda, 
Tokyo 102-0075, Japan}
\author{K. Takenaka}
\affiliation{Department of Crystalline Materials Science, Nagoya University, 
Furo-cho, Chikusa-ku, Nagoya 464-8603, Japan}
\affiliation{TRIP, Japan Science and Technology Agency (JST), Chiyoda, 
Tokyo 102-0075, Japan}
\author{H. Ikuta}
\affiliation{Department of Crystalline Materials Science, Nagoya University, 
Furo-cho, Chikusa-ku, Nagoya 464-8603, Japan}
\affiliation{TRIP, Japan Science and Technology Agency (JST), Chiyoda, 
Tokyo 102-0075, Japan}
\author{Y. Takayanagi}
\affiliation{Department of Physics, Faculty of Science, Nagoya University, 
Furo-cho, Chikusa-ku, Nagoya 464-8602, Japan}
\author{N. Hayamizu}
\affiliation{Department of Physics, Faculty of Science, Nagoya University, 
Furo-cho, Chikusa-ku, Nagoya 464-8602, Japan}
\author{Y. Sone}
\affiliation{Department of Physics, Faculty of Science, Nagoya University, 
Furo-cho, Chikusa-ku, Nagoya 464-8602, Japan}
\date{\today}

\begin{abstract}
Two magnon excitations and the nodal spin density wave (SDW) gap were 
observed in BaFe$_2$As$_2$ by Raman scattering.  
Below the SDW transition temperature ($T_{\rm SDW}$) nodal SDW gap opens 
together with new excitations in reconstructed electronic states.  
The two-magnon peak remains above $T_{\rm SDW}$ and moreover the energy 
increases a little.  
The change from the long-range ordered state to the short-range correlated 
state is compared to the cuprate superconductors.  
\end{abstract}

\maketitle
Iron pnictides form a new family of superconductors with the 
transition temperature up to $T_{\rm c}=55$ K \cite{Kamihara,Ren}.  
The superconductivity emerges as the spin density wave (SDW) is reduced 
by substituting an element or pressurizing.  
The stripe type spin order in the SDW state and the superconductivity are 
generally assumed to originate from the nesting of the hole Fermi surface (FS)
at $\Gamma$ and the electron FS at $M$ \cite{Mazin,Dong,Kuroki,Ding,Terashima}.  
The first issue is the change of the electronic states at the SDW transition, 
especially the SDW gap.  
If the SDW arises from the instability of itinerant electrons for the nesting, 
a gap opens at the FS.  
Ran {\it et al}. \cite {Ran} predicted that the gap is not a full gap but 
a nodal gap.  
Infrared spectroscopy reported the SDW gap \cite{Hu}.  
The present electronic Raman scattering disclosed opening of the gap 
in the $(ab)$ and $(xx)$ spectra, but not in $(aa)$ and $(xy)$, indicating 
the nodal gap, were $(\hat E_{\rm i} \hat E_{\rm s})$ 
denotes that the electric fields of incident and scattered light are parallel 
to $\hat E_{\rm i}$ and $\hat E_{\rm s}$.   
It is consistent with the very recent observation of the anisotropic gap 
observed in ARPES \cite{Hsieh,YiSDW}.
The second issue is what kinds of exchange interactions work in the SDW state 
and what kinds of magnetic correlations remain above the SDW transition 
temperature ($T_{\rm SDW}$).  
The two-magnon peak is observed at 2200 cm$^{-1}$ in consistent with neutron 
scattering \cite{Ewings,Zhao,McQueeney,Matan,Zhao2}. 
In order to explain the magnetic excitation spectra in metal a carrier hopping 
induced mechanism is proposed and compared to the cuprate superconductors.   

BaFe$_2$As$_2$ undergoes the SDW state below 
$T_{\rm SDW}=137$ K \cite{Rotter,Huang}.  
The crystal structure changes from tetragonal $(I4/mmm)$ to orthorhombic 
$(Fmmm)$ \cite{Rotter,Huang}.  
The magnetic order in the SDW state is a stripe type in which nearest-neighbor 
spins are antiparallel in the $x$ direction and parallel 
in the $y$ direction \cite{Cruz,Ewings,Zhao,McQueeney,Matan}.  

Single crystals of BaFe$_2$As$_2$ were grown by the self-flux method.  
Raman spectra were measured on the fresh cleaved surfaces in a quasi-back 
scattering configuration using 5145 \AA \ laser light.  
The crystallographic axes of the tetragonal structure are $a$ and $b$ and 
the bisecting directions are $x$ and $y$.  
Raman active phonon modes are $1A_{\rm 1g}+1B_{\rm 1g}+2E_{\rm g}$ in the 
tetragonal structure \cite{Litvinchuk,Rahlenbeck,Choi,Chauviere}.  
The $A_{\rm 1g}$, $B_{\rm 1g}$, and $E_{\rm g}$ modes changes into the 
$A_{\rm 1g}$, $B_{\rm 1g}$, and $B_{\rm 2g}+B_{\rm 3g}$ modes 
in the orthorhombic structure, respectively.  
The $(aa)$ spectra allow the $A_{\rm 1g}+B_{\rm 1g}$ ($A_{\rm 1g}+B_{\rm 1g}$) 
modes, $(ab)$ no ($A_{\rm 1g}$) mode, 
$(xx)$ $A_{\rm 1g}+B_{\rm 2g}$ ($A_{\rm 1g}$), and $(xy)$ $B_{\rm 1g}$ 
($B_{\rm 1g}$) in the tetragonal (orthorhombic) structure.  
It should be noted that the crystallographic axes rotate by $45^{\circ}$ 
as the structure changes from tetragonal to orthorhombic.
The Raman system was calibrated so that the intensity is proportional 
to ${\partial ^2 S}/{\partial \omega \partial \Omega}$.

\begin{figure}
\begin{center}
\includegraphics[trim=0mm 0mm 0mm 0mm, width=8.5cm]{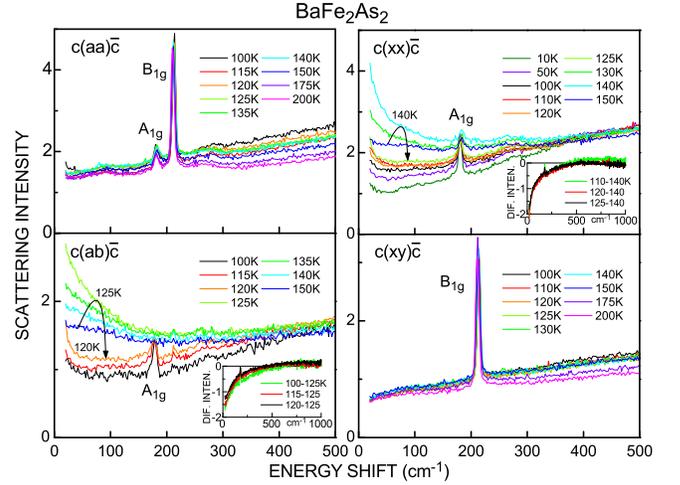}
\caption{(color online) 
Temperature dependent low-energy Raman spectra in BaFe$_2$As$_2$.  
The $c(ab)\bar c$ and $c(xx)\bar c$ spectra show the SDW gap, but the 
$c(aa)\bar c$ and $c(xy)\bar c$ spectra do not show the gap.  
The insets are the differential spectra between below and above $T_{\rm SDW}$.
}
\label{fig3}
\end{center}
\end{figure}
The SDW gap is directly observed in the low-energy spectra in Fig. 1.  
In the $(ab)$ and $(xx)$ spectra the low energy part below 200 cm$^{-1}$ 
increases as temperature decreases from 150 K and then abruptly decreases 
from 125 K to 120 K in $(ab)$ and from 140 K to 125 K in $(xx)$.  
The differential spectra are shown in the insets.  
The small temperature difference is due to the first-order phase transition.  
The first increases of the low-energy intensity is caused by the increase of 
the low-energy magnetic fluctuation \cite{Matan}.  
The following abrupt decrease is due to the opening of the SDW gap.  
It should be noted that the full gap does not open as known from the metallic 
properties in the SDW state.  
Ran {\it et al}. \cite{Ran} calculated that the hole pocket at $\Gamma$ and 
the electron pocket folded into $\Gamma$ make a pseudo-gap with two 
connecting points by the Dirac nodes.  
The connecting points are a little above the Fermi level and two small hole 
pockets are made.  
Similarly the electron pocket at $(0, \pi)$ and the hole pocket folded into 
$(0, \pi)$ make a pseudo-gap and two small electron pockets.  
The gap energy is about 400 cm$^{-1}$, when it is assigned to the starting 
point of the deviation between the spectra above and below $T_{\rm SDW}$.  
This is the lowest-energy pseudo-gap.  
The gap structure is not observed in the $(aa)$ and $(xy)$ spectra.  
Therefore the gap symmetry is $B_{\rm 2g}$.  
It is the same as the superconducting gap in BaFe$_{2-x}$Co$_x$As$_2$ \cite{Muschler,Sugai}.  
The hole pocket gives the largest contribution to the electronic scattering 
intensity in the $(aa)$ spectra, when it is calculated from the 
$|\partial ^2E/\partial k_i \partial k_s |^2$, where $k_i$ and $k_s$ are 
the wave vectors in the incident and scattered polarization directions.  
The reason that the gap does not appear in $(aa)$ is the multi-orbital effect.  
The total symmetry of the paired state is given by the symmetries in the 
orbital combination and the momentum space.  
The observed gap symmetry is gives by the $B_{\rm 2g}$ symmetry in the orbital 
combination and the $A_{\rm 1g}$ symmetry in the momentum space.  
The details are presented separately \cite{Sugai}.

The sharp peaks at 181 ($A_{\rm 1g}$) and 215 cm$^{-1}$ ($B_{\rm 1g}$) 
are phonon peaks \cite{Litvinchuk,Rahlenbeck,Choi,Chauviere}.  
Below $T_{\rm SDW}$ the 187 cm$^{-1}$ $A_{\rm 1g}$ phonon peak 
appears in the $(ab)$ spectra.  
In this polarization configuration the forbidden $A_{\rm 1g}$ phonon 
in the tetragonal structure becomes allowed in the orthorhombic structure, 
but the intensity is expected to be small because it is proportional to 
$|R_{11}-R_{22}|^2$ with $R_{11}\approx R_{22}$ using the Raman tensor of the 
$A_{\rm 1g}$ phonon.  
The intensities in the $(aa)$ and $(xx)$ spectra are proportional to 
$\frac{1}{4} |R_{11}+R_{22}|^2$ and $|R_{11}|^2$ (or $|R_{22}|^2$), respectively.  
Both intensities are expected to be close to each other as 
$R_{11}\approx R_{22}$ and the twinning are expected.  
Instead at 10 K the intensities in the $(xx)$ and $(ab)$ spectra in which 
the SDW gap opens are $1.5\sim 2$ times as large as the intensity 
in $(aa)$ in which the gap closes.   
This phonon is the mode in which As atom moves in the $c$ 
direction \cite{Litvinchuk}.  
It has large magneto-phonon interaction, because the Fe-As-Fe angle is very 
sensitive to the Fe-Fe exchange interaction energy \cite{Yildirim,Yin,Kuroki2}.  
The detailed mechanism of the enhancement is still an open question.  
The enhancement of the $A_{\rm 1g}$ phonon below $T_{\rm SDW}$ was reported 
in CaFe$_2$As$_2$ \cite{Choi} and Ba(Fe$_{1-x}$Co$_x$)$_2$As$_2$ \cite{Chauviere}.  
Similar enhancement of the infrared active phonon was reported in 
BaFe$_2$As$_2$ \cite{Akrap}

\begin{figure}
\begin{center}
\includegraphics[trim=0mm 0mm 0mm 0mm, width=7.5cm]{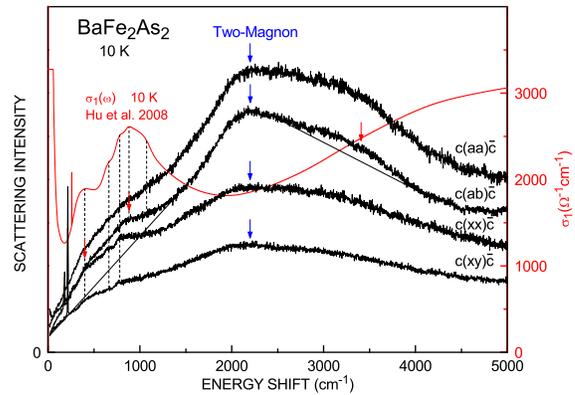}
\caption{(color online) 
Comparison between Raman spectra and optical conductivity at 10 K \cite{Hu}.  
}
\label{fig2}
\end{center}
\end{figure}

\begin{figure*}
\begin{center}
\includegraphics[trim=0mm 0mm 0mm 0mm, width=14cm]{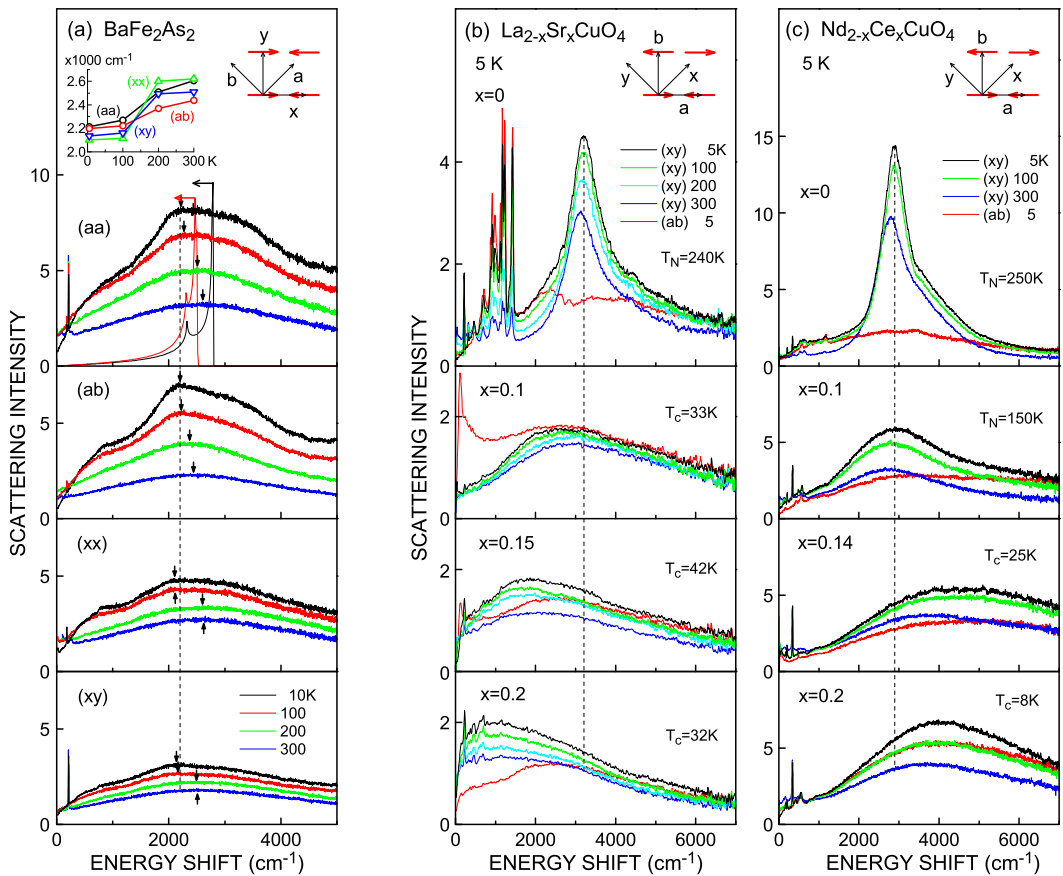}%
\caption{(color online) 
(a) Temperature dependent polarized Raman spectra in BaFe$_2$As$_2$.  
The two-magnon peak is shown by the dashed line.  
The densities of states for the independent two spin wave excitations are 
shown with $SJ_{1a}=36$, $SJ_{1b}=-7.2$, and $SJ_2=18$ meV \cite{Ewings} 
(black) and $SJ_{1a}=43$, $SJ_{1b}=-3.1$, and $SJ_2=14.3$ meV \cite{Han} 
(red) in the (aa) spectra.  
The two-magnon peak energy is reduced by the magnon-magnon interaction 
energy as shown by the left-pointing arrows.  
The inset is the two-magnon peak energy.  
(b) Two-magnon Raman spectra in hole-doped LSCO and (c) electron-doped NCCO.  
Two-magnon scattering in the insulating phase is allowed in the $(xy)$ and 
$(aa)$ spectra, but not in $(ab)$ and $(xx)$.  
The low-energy peaks in LSCO with $x=0.1$ and 0.15 in the $(ab)$ spectra 
are superconducting coherent peaks.  
}
\label{fig1}
\end{center}
\end{figure*}
Figure 2 shows the 10 K spectra at various polarization configurations and 
the optical conductivity at 10 K obtained by Hu et al. \cite{Hu}. 
The temperature dependence of the Raman spectra are shown in Fig. 3(a).
The humps below 1000 cm$^{-1}$ and 2800-3800 cm$^{-1}$ disappear above $T_{\rm SDW}$.  
Therefore these humps are made by the electronic transition in 
the reconstructed electronic states created by 
the folding of the electron pockets and the hole pockets below $T_{\rm SDW}$.  
The new broad humps at 400, 900, and 3500 cm$^{-1}$ correspond to the 
360, 890, and 5000 cm$^{-1}$ humps observed in the optical conductivity \cite{Hu}
The 662 and 780 peaks in the $(xx)$ and $(xy)$ spectra and the 882 and 
1072 peaks in the $(aa)$ and $(ab)$ spectra correspond to the fine structure 
in the optical conductivity spectra \cite{Hu}.  
The energy difference 118 cm$^{-1}$ in the former set is close to the 
$E_{\rm g}$ phonon energy 117 cm$^{-1}$ and the energy difference 190 cm$^{-1}$ 
in the latter set is close to the $A_{\rm 1g}$ phonon energy 182 
cm$^{-1}$ \cite{Rahlenbeck}.  
Very recently Yi et al. \cite{YiSDW} reported the ARPES results that the 
electronic structures are significantly reconstructed in the SDW state and 
cannot be described in the simple folding scenario.  

The broad 2200 cm$^{-1}$ (at 10 K) peak is the two-magnon scattering peak.  
Till now two-magnon scattering is interpreted by the Coulomb interaction 
induced $S^+ S^-$ excitations on the spin waves at $k$ and $-k$.  
This process expresses two-magnon scattering in the antiferromagnetic (AF) 
insulator.  
However, it is noted that BaFe$_2$As$_2$ is metal even in the SDW state.  
We first discuss the two-magnon peak by the Coulomb interaction induced 
mechanism and then discuss the new carrier induced mechanism 
by comparing with the cuprate superconductors.

The exchange interaction energies are very different whether the calculation 
starts from the long-range AF stripe spin structure or from 
the short-range superexchange interaction.  
In the former case the exchange interaction energies are obtained from 
$J_{ij}(R)=-\partial ^2E/\partial \theta _i(0) \partial \theta _j(R)$, 
where $E$ is the total energy and $\theta _j(R)$ is the angle of the moment 
of the $j$th spin \cite{Yin,Han}.  
The calculated energies are AF $SJ_{1x}=43$, ferromagnetic 
$SJ_{1y}=-3.1$, and AF diagonal $SJ_2=14.3$ meV \cite{Han}.  
The low-energy spin wave in the SDW state was observed by neutron scattering 
in BaFe$_2$As$_2$ \cite{Ewings,Matan}, SrFe$_2$As$_2$ \cite{Zhao} 
and CaFe$_2$As$_2$ \cite{McQueeney}.  
Ewings {\it et al}. \cite{Ewings} fitted the velocity at 7 K in BaFe$_2$As$_2$ 
by the long-range exchange interaction model with $SJ_{1x}=36$, 
$SJ_{1y}=-7.2$, and $SJ_2=18$ meV.  
Recent high energy neutron scattering experiment showed that the entire 
spin wave dispersion of CaFe$_2$As$_2$ ($T_{\rm N}\sim 170$ K) at 10 K is 
expressed by the long-range model \cite{Zhao2}.  
On the other hand the calculation of the local superexchange interaction 
gives AF exchange interaction energies for all directions 
reflecting the equivalent $x$ and $y$ directions.  
The stripe spin order is stable if $J_{1x}=J_{1y}<2J_2$, otherwise 
the checkerboard spin order is stable \cite{Yildirim,Si}.  
The two-magnon peak is observed even above $T_{\rm SDW}$ in Fig. 3(a), 
indicating that the short-range spin correlation remains.  
The $J_{1x}=J_{1y}$ type of exchange interactions may appear above 
$T_{\rm SDW}$.

The densities of states of the spin waves calculated using the exchange 
interaction energies of the first principle calculation (red) and the neutron 
scattering (black) are shows in the $(aa)$ spectra of Fig. 3(a).  
The energies are doubled to present independent two magnon excitations.  
The observed two-magnon peak energy is reduced by the magnon-magnon interaction 
energy of about $J_{1x}$ as shown by the left-pointing arrow assuming $S=1$.  
The calculated two-magnon energy is close to the experimentally obtained 
two-magnon peak energy.

The two-magnon scattering Hamiltonian is \cite{Parkinson}
\begin{equation}
H'=A\sum_{i,j} (\hat E_{\rm s}\cdot \hat \rho_{ij})(\hat E_{\rm i}\cdot \hat 
\rho_{ij})({\bf S}_i \cdot {\bf S}_j),
\end{equation}
where $\hat \rho_{ij}$ is the unit vector connecting near spins on the 
different spin sublattices.  
From this Hamiltonian two-magnon scattering is active in all polarization 
configurations in contrast to the cuprate in which the active polarizations are 
$(aa)$ and $(xy)$.  
The difference comes from the large diagonal exchange interaction energy in 
BaFe$_2$As$_2$.

The very broad two-magnon peak in BaFe$_2$As$_2$ is caused by itinerant 
carriers.  
The hole density is 0.081 and the electron density is 0.069 
per iron atom at 140 K \cite{Yi}.  
Figure 3(b) and (c) show the doping dependence in hole-doped 
La$_{2-x}$Sr$_x$CuO$_4$ (LSCO) and electron-doped 
Nd$_{2-x}$Ce$_x$CuO$_4$ (NCCO).  
The two-magnon peak energy in the $(xy)$ spectra of LSCO decreases with 
increasing the hole density, whereas it abruptly increases at the 
insulator-metal transition in NCCO.  
It is noted that the two-magnon peak is observed above the N\'eel temperature 
($T_{\rm N}$) and in the superconducting state, because the short-range 
correlation remains.  
Till now the effect of carriers on two-magnon scattering is treated by the 
Coulomb interaction induced insulating two-magnon scattering process with the 
reduced AF correlation length. 
In this model the two-magnon peak decreases in energy and intensity 
with broadening as the carrier density increases.  
The increase of the two-magnon peak energy in the metallic phase of NCCO 
cannot be interpreted.  
We propose a new model that electronic scattering gives the magnetic 
excitation component.

A hole or an electron traveling in the antiferromagnetically ordered spin sea 
leaves behind an changed spin trace, because the carrier spin which hops 
to the neighboring site is limited by the Pauli principle.  
The electron Green function representing this state has a self-energy of 
spin excitations.  
The electron spectral function that is the imaginary part of the retarded 
Green function has the coherent component near the original electronic energy 
and the incoherent component near the spin excitation energy at a fixed $k$.  
It is calculated by the string model \cite{Manousakis}.  
The $\Delta k\approx 0$ transition in the electronic Raman scattering detects 
the magnetic excitations through the incoherent part.  
The minimum spin excitation is two spin changes of $\Delta S_z=1$ and 
$\Delta S_z=-1$ because of the spin conservation with light.  
The number of changed spins increases and the energy increases as the mean 
free path of the carrier increases.  
The mean free path in NCCO is ten times longer than that in LSCO.  
It causes the high energy shift of the magnetic excitation peak in NCCO.  
In case of BaFe$_2$As$_2$ the observed magnetic excitation peak energy is nearly 
the same as the calculated two-magnon energy.  
It indicates that the mean free path is not as long as NCCO and the magnetic 
correlation length is not as short as LSCO.

The scattering intensity above the maximum of the independent two spin wave 
excitations comes from the higher order multi-magnon excitations.  
The higher-order component is much larger in the carrier-induced case than 
the Coulomb interaction-induced case.  
The large higher-order component in BaFe$_2$As$_2$ is caused by the 
carrier-induced process.  
The mean free path decreases from $T_{\rm SDW}$ to 300 K by 15 \%, when it is 
estimated from the in-plane resistivity.  
Hence the carrier-induced process does not contribute to increase the higher 
order multi-magnon scattering.  
The 2200 cm$^{-1}$ two-magnon peak shifts to $2400 \sim 2600$ cm$^{-1}$ as 
temperature increases from $T_{\rm SDW}$ to 300 K (inset of Fig. 3(a)).  
The optical conductivity in the two-magnon energy region and the plasma 
frequency does not show appreciable change above $T_{\rm SDW}$ \cite{Hu}.  
Therefore the high energy shift may be induced by the inclusion of 
the short-range spin correlation, AF in both $x$ and $y$.

The present Raman scattering experiment disclosed the anisotropic SDW gap 
and two-magnon scattering.  
Two-magnon peak survives far above $T_{\rm SDW}$ indicating the 
short-range spin correlation remains.  
The two-magnon scattering process in metal was discussed in comparison with 
the cuprate superconductors.  

This work was supported by Transformative 
Research Project on Iron Pnictides (TRIP), Japan Science and Technology 
Agency (JST).

\end{document}